\pdfoutput=1
\documentclass[aps,prx,twocolumn,groupedaddress,nofootinbib]{revtex4-2}

\usepackage{amsmath,amssymb,bm}
\usepackage{graphicx}
\usepackage{xcolor}
\usepackage{hyperref}
\usepackage{booktabs}
\usepackage{siunitx}

\begin{document}

\title{Inter-branch message transfer on superconducting quantum processors: a multi-architecture benchmark}

\author{Cameron V. Cogburn}
\email{cogbuc@rpi.edu}
\affiliation{Future of Computing Institute, Rensselaer Polytechnic Institute, Troy, NY 12180, USA}

\date{January 2026}

\begin{abstract}
We treat inter-branch message transfer in a Wigner's-friend circuit as a practical benchmark for near-term superconducting quantum processors. Implementing Violaris' unitary message-transfer primitive, we compare performance across IBM Eagle, Nighthawk, and Heron (r2/r3) processors for message sizes up to $n=32$, without error mitigation. We study three message families---sparse (one-hot), half-weight, and dense---and measure conditional string success $p_{\mathrm{all}}=\Pr(P=\mu\mid R=0)$, memory erasure after uncomputation, and correlation diagnostics (branch contrast and bitwise mutual information). The sparse family compiles to essentially constant two-qubit depth, yielding a depth-controlled probe of device noise: at $n=32$ we observe $p_{\mathrm{all}}$ spanning $\approx0.07$ to $\approx0.68$ across backends. In contrast, half and dense messages incur rapidly growing routing overhead, and transpiler-seed variability becomes a practical limitation near the coherence frontier. We further report an amplitude sweep (no-amplification test) and a divergence ``cousins'' sweep that quantifies degradation with branch-conditioned complexity. All data and figure-generation scripts are released.
\end{abstract}

\maketitle

\section{Introduction}

It is often thought that observers in distinct Everett branches cannot communicate without violating linear quantum dynamics. Violaris challenges this intuition within standard unitary quantum mechanics with an explicit, unitary circuit-level protocol in a Wigner's-friend setting where an observer (Wigner) has quantum control over another observer (the friend) \cite{Violaris2026Interbranch}. In the ideal circuit model, a classical message $\mu$ written by the friend in the $R=1$ branch can be recovered on a ``paper'' register in the $R=0$ branch, while the friend's memory is uncomputed to preserve global unitarity.

In Ref.~\cite{Altman2026Witness}, Altman implements and benchmarks an instance on specific superconducting quantum hardware (\texttt{ibm\_fez}), emphasizing coherence-sensitive witnesses and the role of error-mitigation constraints. Our focus here is complementary and operational: we treat inter-branch message transfer as a parameterized circuit family and quantify how raw transfer and erasure performance depends on message size, message structure (Hamming weight), device family, and compilation choices (e.g., transpiler seeds). Because the sparse message family compiles to nearly constant two-qubit depth, it serves as a depth-controlled probe of device noise. In contrast, the half and dense message families expose routing overhead and connectivity limits.

Wigner's-friend scenarios have recently been revisited from multiple angles, including no-go theorems for observer-independent facts and single-world consistency, and photonic experiments that realize extended Wigner's-friend networks~\cite{Wigner1961,Brukner2018NoGo,FrauchigerRenner2018,Bong2020NoGo,Proietti2019}. The present work does not implement those multi-observer nonlocal tests. Instead, we focus on Violaris' unitary “message-transfer” primitive as a standalone circuit component, and ask what can be achieved on current superconducting devices without mitigation.

\paragraph{Contributions.} In this work we provide:
\begin{itemize}
    \item a reproducible Qiskit implementation of the inter-branch message-transfer circuit family and key controls, with scaling up to $n=32$ message bits;
    \item a multi-architecture benchmark across IBM Eagle (\texttt{ibm\_rensselaer}), Nighthawk (\texttt{ibm\_miami}), and two Heron revisions (\texttt{ibm\_fez} and \texttt{ibm\_boston});
    \item depth-controlled device-to-device comparisons via the nearly constant compiled two-qubit depth sparse message family, contrasted with routing-limited half and dense message families;
    \item an empirical characterization of transpiler-seed variability near the coherence frontier; and
    \item two protocol stress tests: an amplitude sweep (no-amplification corollary) and a “cousins” divergence sweep (twins vs cousins).
\end{itemize}

Finally, we emphasize that the circuits are standard unitary constructions; our goal is not to discriminate among interpretations of quantum mechanics, but to provide an experimentally grounded, reproducible benchmark suite for this circuit primitive on current superconducting architectures.

\section{Protocol, message families, and metrics}

Following~\cite{Violaris2026Interbranch}, the global Hilbert space is decomposed into the following registers: the measured qubit $Q$, a branch label $R$, the friend's state qubit $F$, the friend's memory register $M$ (an $n$-qubit register), and the paper register $P$ (also an $n$-qubit register). The protocol is as follows. The friend measures $Q$ into $F$ (modelled by a CNOT), the branch label $R$ records the friend's branch (CNOT from $F$ to $R$), and the friend in the $R=1$ branch writes the classical message $\mu$ into $M$ and then onto $P$. Wigner then uncomputes the memory $M$ using message-independent operations (CNOT from $P$ to $M$), and applies a global unitary partial branch swap
\begin{equation}
U_{\rightleftharpoons} = X_Q \otimes X_R \otimes X_F,
\end{equation}
which exchanges branch-defining degrees of freedom but leaves $M$ and $P$ unchanged. This relocates the paper record $\mu$ from the $R=1$ branch to the $R=0$ branch.

\subsection{Message families and operational metrics}\label{sec:message-families}

The above construction produces a circuit only requiring controlled operations and can be easily implemented on superconducting quantum hardware. To probe how compilation and noise scale on the hardware, we define three message families:
\begin{itemize}
  \item \textbf{Sparse:} $\mu=10\cdots0$, so the compiled two-qubit depth of the write and uncompute is essentially constant in message size $n$.
  \item \textbf{Half:} $\mu$ is chosen with Hamming weight $k=\lfloor n/2\rfloor$ by selecting $k$ bit positions uniformly at random and setting those bits to 1 (the others to 0)\footnote{For each $n$ we sample several such $\mu$ and report mean $\pm$ standard deviation across instances. All $\mu$ strings are recorded in the data files for reproducibility.}.
  \item \textbf{Dense:} $\mu=11\cdots1$, which typically yields the largest entangling depth after compilation.
\end{itemize}

We then report on three primary observables:
\begin{itemize}
    \item \textbf{Bitwise transfer:}
    \[
    p_{\mathrm{bit}}=\frac{1}{n}\sum_i \Pr(P_i=\mu_i\mid R=0).
    \]
  \item \textbf{String transfer:}
    $p_{\mathrm{all}}=\Pr(P=\mu\mid R=0)$.
  \item \textbf{Memory erasure:}
    \[
    p_{\mathrm{erase}}=\Pr(M=0\cdots 0\mid R=1) 
    \] 
    after uncomputation.
\end{itemize}
We also track two auxiliary diagnostics to distinguish the strict, all-bits-correct success metrics above from information transfer that can remain nonzero even when the all-bits-correct events becomes rare at large circuit depths:
\begin{itemize}
    \item \textbf{Branch contrast:} $\Delta$, the mean difference over active bits ($\mu_i=1$) between the probability that $P_i=1$ in the $R=0$ branch and the probability that $P_i=1$ in the $R=1$ branch. Positive $\Delta$ indicates the message ``lives'' in $R=0$.
    \item \textbf{Mutual information:} the bitwise mutual information $I(R;P_i)$, averaged over active bits, quantifying residual correlations between the branch label and the transferred paper bit values.
\end{itemize}

\section{Experimental methods}

We implement the circuits in Qiskit~\cite{JavadiAbhari2024Qiskit,Qiskit} and execute them on IBM superconducting quantum processors via IBM Quantum cloud services~\cite{IBMQuantum} using the default transpiler and backend-specific coupling constraints. 
We benchmark four different backends spanning three processor families:
\texttt{ibm\_rensselaer} (Eagle),
\texttt{ibm\_miami} (Nighthawk),
\texttt{ibm\_fez} (Heron r2),
and \texttt{ibm\_boston} (Heron r3).
Unless otherwise noted, experiments use 4096 shots per circuit and transpiler optimization level 3. Error bars denote the standard deviation across compiled instances (repeated executions and/or transpiler-seed sweeps). All results are without error mitigation.

\section{Results}

\subsection{Pilot benchmark and controls}

We begin by implementing the $n=1$ protocol and two controls. The first control, \texttt{no\_swap}, omits the partial branch swap (so the message remains in $R=1$). The second control, \texttt{no\_uncompute}, omits memory uncomputation, so the sender retains a memory record of $\mu$. The latter control directly probes Corollary~1 of~\cite{Violaris2026Interbranch}.

\begin{figure}
  \centering
  \includegraphics[width=\linewidth]{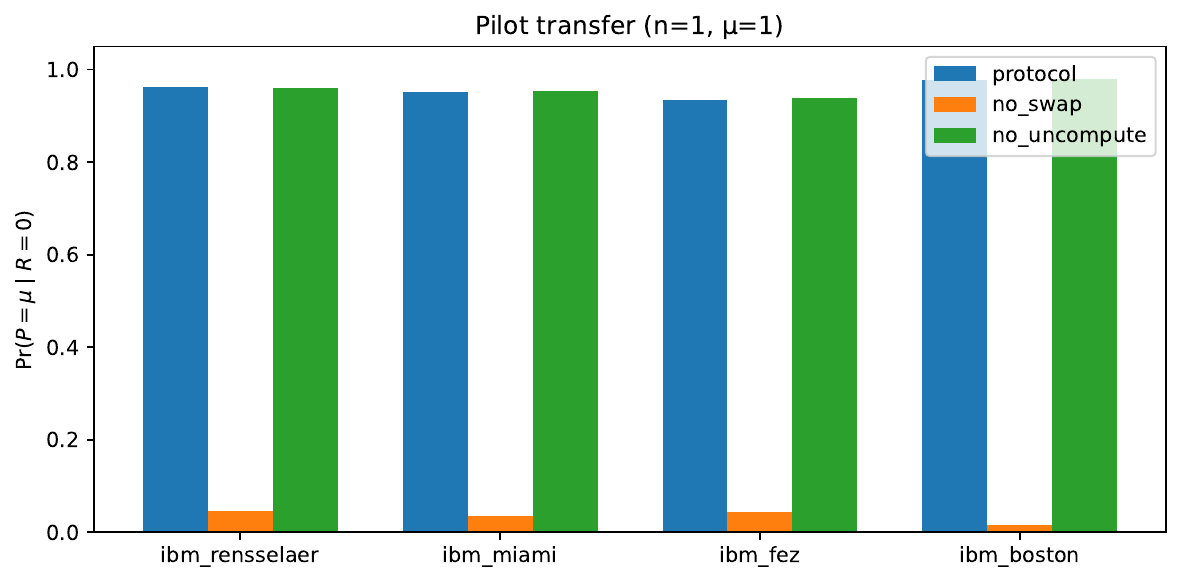}
  \caption{Pilot benchmark for string-level transfer success at $n=1$ for the protocol and controls.}
  \label{fig:pilot-transfer-n1}
\end{figure}

\begin{figure}
  \centering
  \includegraphics[width=\linewidth]{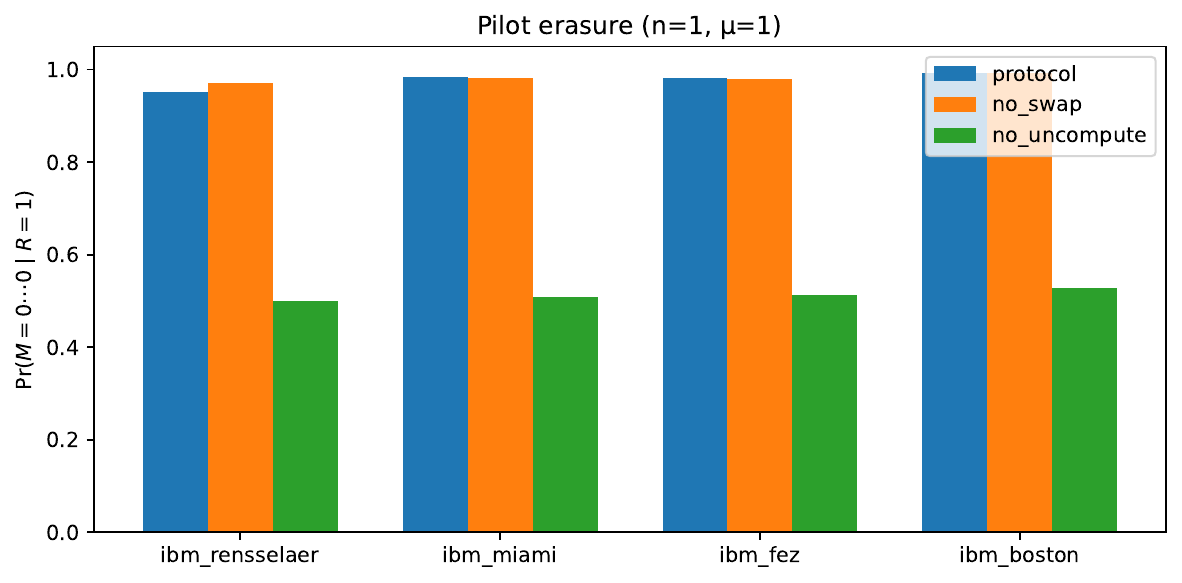}
  \caption{Pilot benchmark for memory erasure at $n=1$ for the protocol and controls.}
  \label{fig:pilot-erasure-n1}
\end{figure}

Figure~\ref{fig:pilot-transfer-n1} reproduces this basic implementation on all backends. This shows the protocol transfers the message with a high probability, while the ``no swap" control fails, as expected. Figure~\ref{fig:pilot-erasure-n1} shows that uncomputation is essential for erasure: the \texttt{no\_uncompute} control preserves transfer but strongly degrades memory erasure.

\subsection{Scaling with message size}

\begin{figure}
  \centering
  \includegraphics[width=\linewidth]{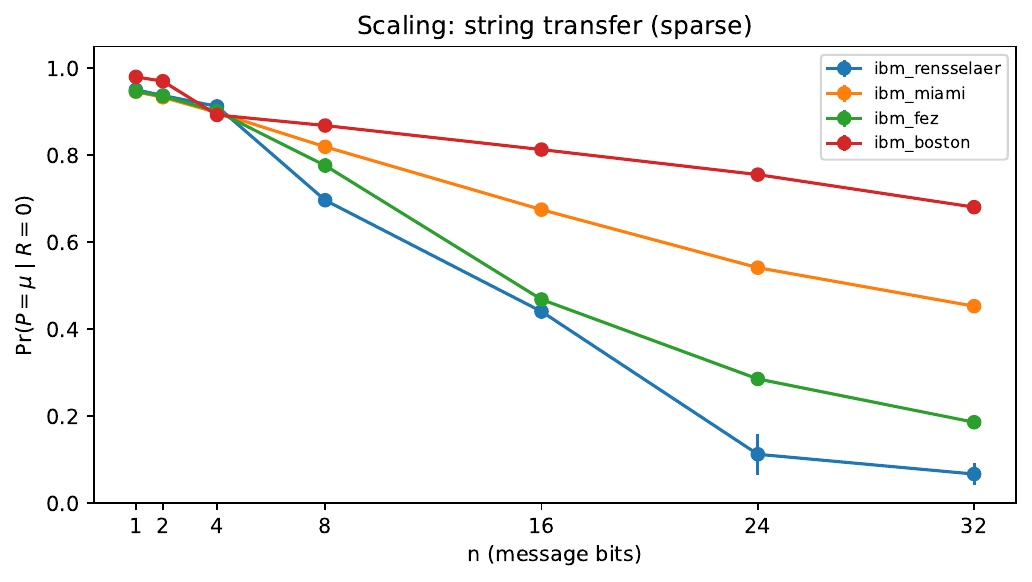}
  \caption{Scaling of string transfer success for sparse messages. The compiled two-qubit depth is essentially constant, so performance differences largely reflect device-level noise.}
  \label{fig:scaling-transfer-sparse}
\end{figure}

\begin{figure}
  \centering
  \includegraphics[width=\linewidth]{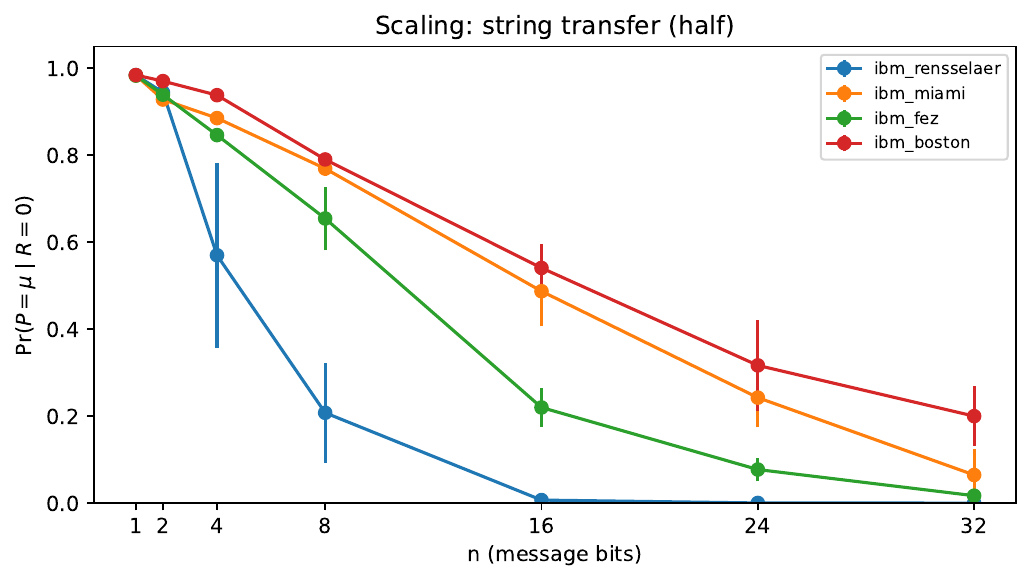}
  \caption{Scaling of string transfer success for half-weight messages.}
  \label{fig:scaling-transfer-half}
\end{figure}

\begin{figure}
  \centering
  \includegraphics[width=\linewidth]{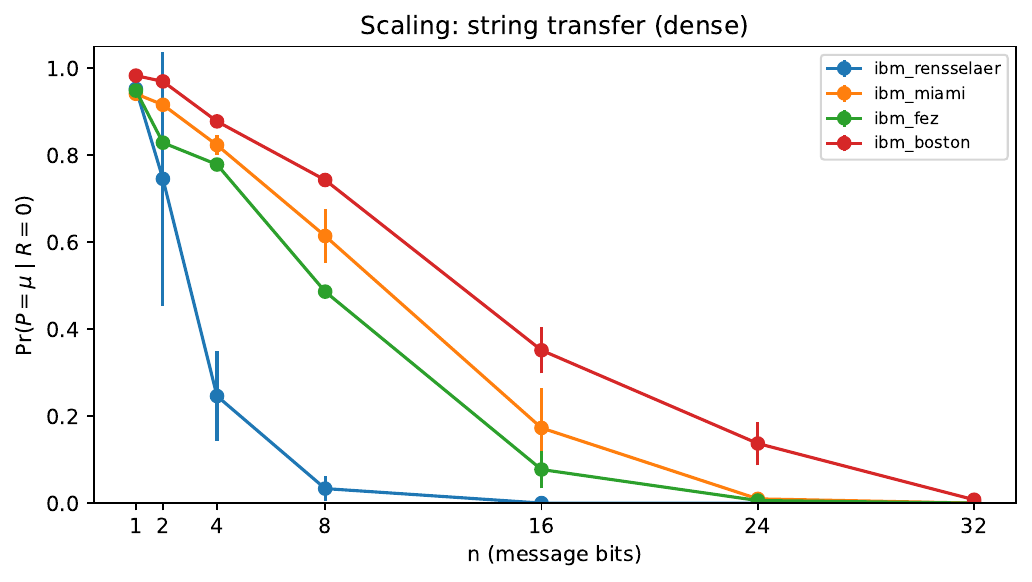}
  \caption{Scaling of string transfer success for dense messages.}
  \label{fig:scaling-transfer-dense}
\end{figure}

Figures~\ref{fig:scaling-transfer-sparse}--\ref{fig:scaling-transfer-dense} show the string transfer probability $p_{\mathrm{all}}$ as a function of message bit size $n$ for the three message families. Recall from their definitions in Sec.~\ref{sec:message-families} that the sparse family isolates device-level performance because the compiled two-qubit depth remains essentially constant, while the half and dense families accumulate increasing circuit depth from routing and entangling.

\begin{figure}
  \centering
  \includegraphics[width=\linewidth]{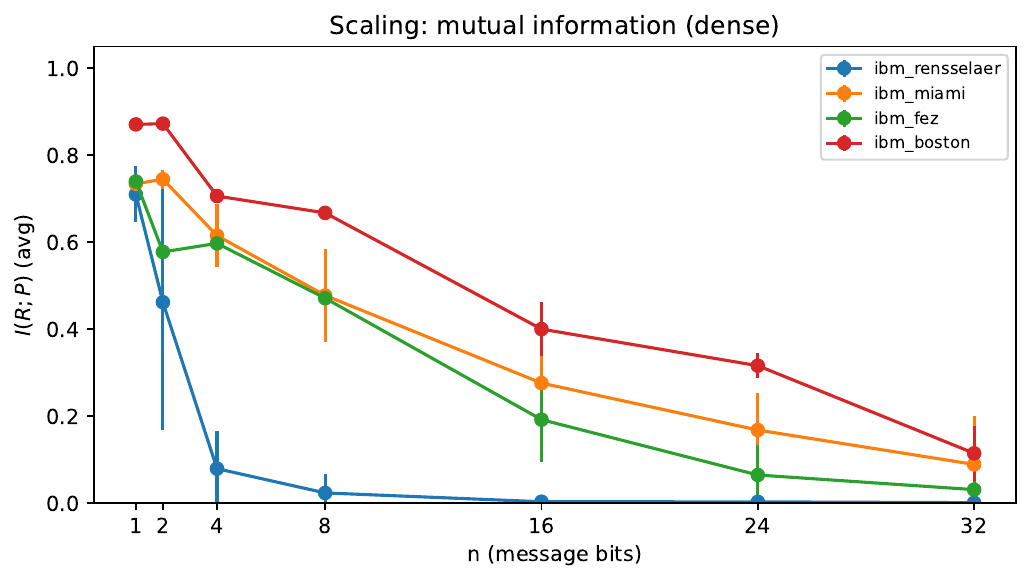}
  \caption{Bitwise mutual information between the final branch label $R$ and the paper bits $P_i$, averaged over active bits ($\mu_i=1$), versus message size $n$ for the dense family. This diagnostic remains nonzero even when the string transfer probability $p_{\mathrm{all}}$ becomes small.}
  \label{fig:scaling-mi-dense}
\end{figure}

Table~\ref{tab:frontier} summarizes a simple frontier statistic: the largest $n$ achieving a mean value of $p_{\mathrm{all}}\ge 0.1$ for each family. Across all families, \texttt{ibm\_boston} achieves the largest frontier bit size, while \texttt{ibm\_rensselaer} is strongly limited for half and dense messages at large $n$. 

This frontier ordering is consistent with two effects captured by our metrics. First, newer devices exhibit lower effective noise at comparable compiled depth. This is seen most clearly in the sparse family where the compiled two-qubit depth is essentially constant. Second, for the half and dense families, the dominant limitation is the growth of compiled two-qubit depth due to routing overhead, which depends on device connectivity and compiler choices.

\begin{table}
\centering
\begin{tabular}{lccc}
\toprule
Backend & Dense & Half & Sparse \\
\midrule
\texttt{ibm\_boston} & 24 & 32 & 32 \\
\texttt{ibm\_fez} & 8 & 16 & 32 \\
\texttt{ibm\_miami} & 16 & 24 & 32 \\
\texttt{ibm\_rensselaer} & 4 & 8 & 24 \\
\bottomrule
\end{tabular}
\caption{Largest message size $n$ with mean string-level transfer success $\Pr(P=\mu\mid R=0)\ge 0.1$ for each family.}
\label{tab:frontier}
\end{table}

To distinguish strictly all-bits-correct string transfer success from weaker but still informative correlations, we also compute the bitwise mutual information $I(R;P_i)$ and average over active bits ($\mu_i=1$). Figure~\ref{fig:scaling-mi-dense} shows that for the dense family $I(R;P_i)$ decays with $n$ as circuit depth grows, but can remain nonzero even when $p_{\mathrm{all}}$ is near zero (e.g., $n\in\{24,32\}$ in Fig.~\ref{fig:scaling-transfer-dense}). This shows residual branch--paper correlations that are not captured by the strict string-success metric.

\subsection{Compilation-seed variability}

\begin{figure}
  \centering
  \includegraphics[width=\linewidth]{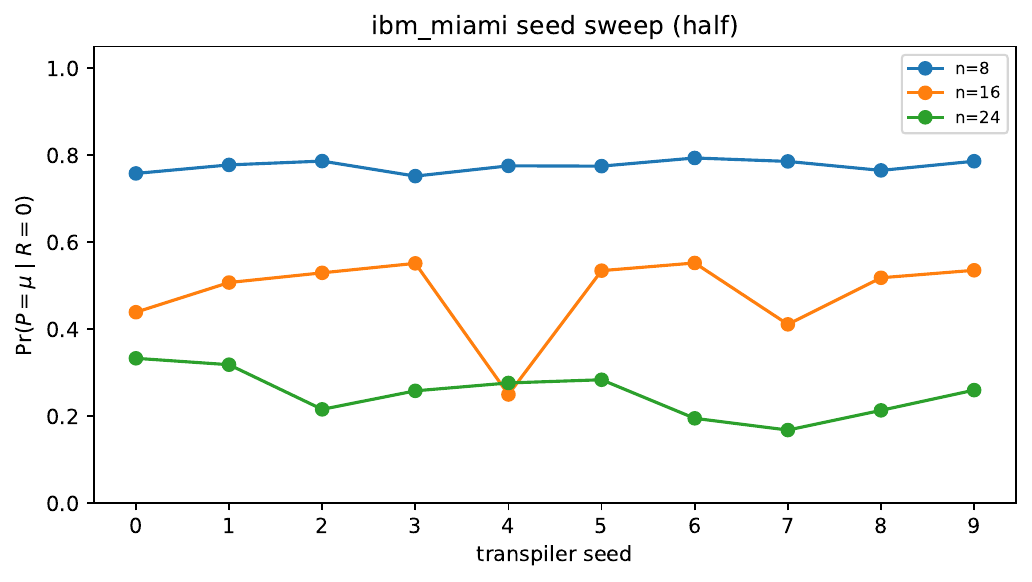}
  \includegraphics[width=\linewidth]{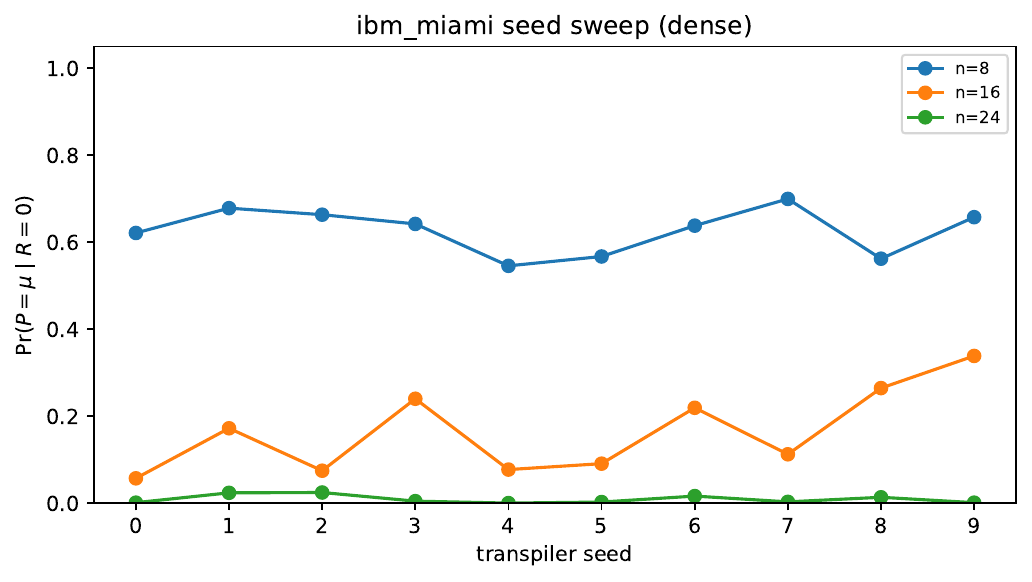}
  \caption{\label{fig:miami-seed-sweep}
    Compilation-seed variability on \texttt{ibm\_miami} for half (top) and dense (bottom) families.}
\end{figure}

Even at fixed $n$ and message family, performance varies across transpiler seeds. Figure~\ref{fig:miami-seed-sweep} shows a seed sweep for \texttt{ibm\_miami} for half and dense messages at $n\in\{8,16,24\}$. This motivates the reporting of error bars and suggests that seed selection can materially affect the largest achievable message size on hardware.

\subsection{``Twins vs cousins'': branch divergence sweep}

\begin{figure}
  \centering
  \includegraphics[width=\linewidth]{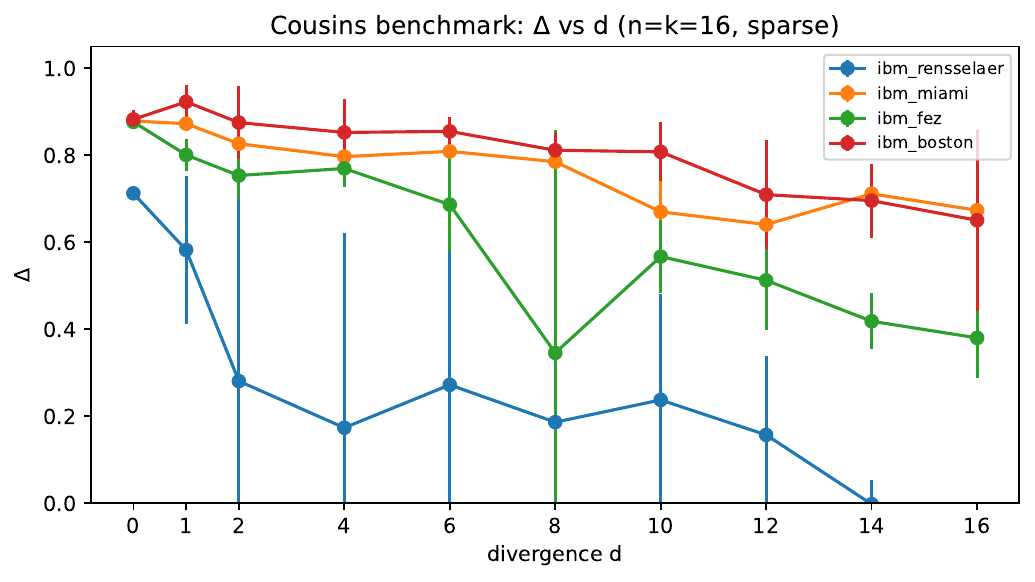}
  \caption{\label{fig:cousins-delta} 
    Cousins benchmark: branch contrast $\Delta$ versus divergence $d$ (mean $\pm$ std across transpiler seeds).}
\end{figure}

Violaris notes that the partial branch swap can become more complex as branch states diverge (``twins'' vs ``cousins'')~\cite{Violaris2026Interbranch}. Therefore, we stress test the protocol with a toy divergence model by adding an unmeasured $k$-qubit ``friend internal state'' register and introducing a Hamming-distance parameter $d$ between branch friend-states. This results in the swap requiring $d$ additional $X$ operations. We fix $k=16$ and sweep over $d\in\{0,1,2,4,6,8,10,12,14,16\}$ for an $n=16$ sparse message. Figure~\ref{fig:cousins-delta} shows the branch contrast $\Delta$ versus divergence $d$. As $d$ increases, the circuit's compiled two-qubit depth grows, with all backends exhibiting degraded branch contrast and transfer.

\subsection{Branch-amplitude sweep}

\begin{figure}
  \centering
  \includegraphics[width=\linewidth]{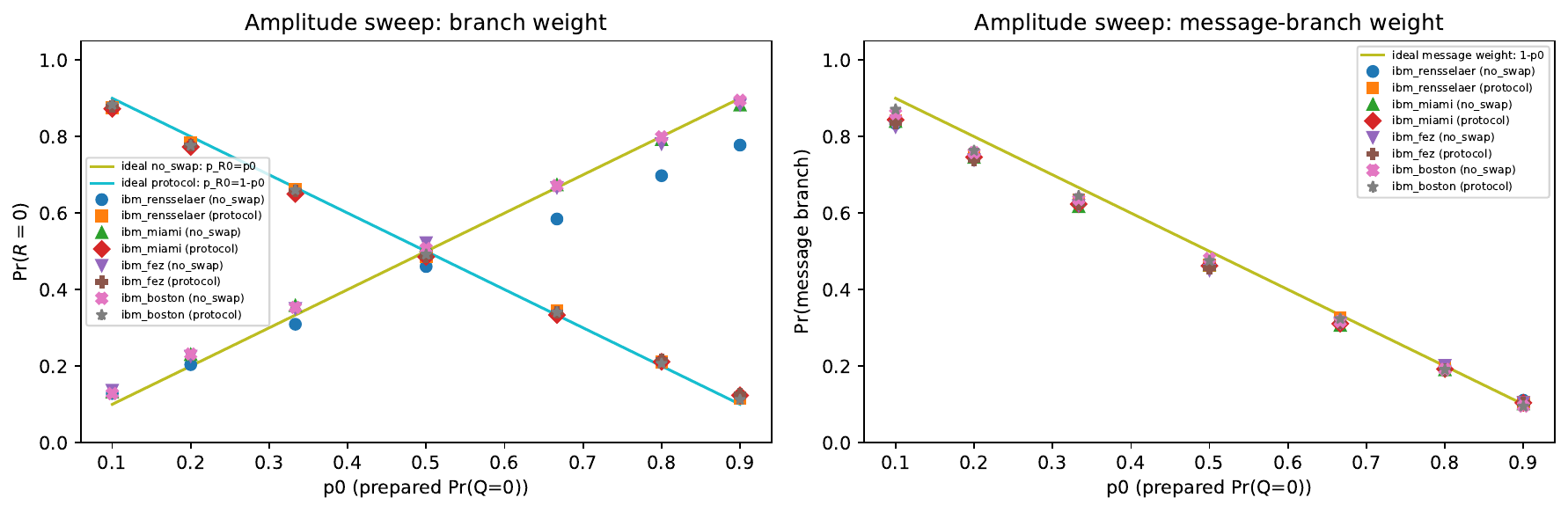}
  \caption{Amplitude sweep. Left: final $\Pr(R=0)$ versus prepared $p_0$, showing the expected branch relabeling. Right: marginal weight of the message branch, which tracks $1-p_0$ and is not amplified.}
  \label{fig:amplitude-sweep}
\end{figure}

In Corollary~2 of \cite{Violaris2026Interbranch}, Violaris notes that message-independent unitaries cannot amplify the total probability weight of the message-bearing branch. We test this by preparing $Q$ so that $\Pr(R=0)=p_0$ and sweeping $p_0$. Figure~\ref{fig:amplitude-sweep} shows (left) how the protocol relabels branch weight ($\Pr(R=0)\approx 1-p_0$ for the protocol, versus $\Pr(R=0)\approx p_0$ for the \texttt{no\_swap} control) and (right) that the marginal message-branch weight tracks the ideal $1-p_0$ and is not amplified.

\section{Discussion}

Our multi-backend results support two practical conclusions. First, for circuits whose compiled two-qubit depth is nearly constant (the sparse message family), hardware-generation differences dominate. At $n=32$ we observe an order-of-magnitude spread in $p_{\mathrm{all}}$ across devices, consistent with the sparse family acting as a depth-controlled probe of effective device noise.

Second, for the half and dense message families, performance is dominated by compilation-induced depth growth from routing overhead. Connectivity and compilation choices therefore matter. The \texttt{ibm\_miami} seed sweep illustrates that transpiler-seed variability can materially shift the achievable message size near the coherence frontier.

Similar to \cite{Altman2026Witness}, we emphasize that these experiments do not discriminate among interpretations of quantum mechanics. The circuits are standard unitary constructions; our contribution is operational: a reproducible benchmark suite for this circuit primitive together with scaling trends and compilation variability across multiple current superconducting architectures.

\section{Conclusion}

We implemented and benchmarked Violaris' inter-branch message-transfer protocol \cite{Violaris2026Interbranch} across four IBM superconducting quantum processors spanning three families, and extended the analysis to scaling, compilation variability, branch-amplitude sweeps, and a cousins divergence benchmark. These results complement \cite{Altman2026Witness} and provide a hardware-level baseline for future devices and mitigation strategies.

\section*{Data and code availability}

All raw results (CSV files) and scripts required to reproduce the figures are included in the reproducibility bundle accompanying this manuscript. Each results file records the IBM Runtime job IDs and execution metadata (backend, instance, transpiler seed). The bundle also includes calibration snapshots to support full provenance. 

\section*{Acknowledgements}
We thank Michael Sofka for bringing Ref.~\cite{Violaris2026Interbranch} to our attention.
We acknowledge the use of IBM Quantum services for this work~\cite{IBMQuantum}. The views expressed are those of the author and do not reflect the official policy or position of IBM or the IBM Quantum team.

\bibliographystyle{apsrev4-2} 
\bibliography{references}

\end{document}